# Ghost diffraction: A spatial statistical approach

MANISHA,[1] VIPIN TIWARI,[2] NANDAN S BISHT,[2, 3] BHARGAB DAS,[4] RAKESH KUMAR SINGH[1,*]

[1]*Laboratory of Information Photonics and Optical Metrology, Department of Physics, Indian Institute of Technology (Banaras Hindu University), Varanasi, 221005, Uttar Pradesh, India.*
[2] *Applied Optics & Spectroscopy Laboratory, Department of Physics, Kumaun University, SSJ Campus Almora 263601, Uttarakhand, India.*
[3] *Department of Physics, Soban Singh Jeena University, Almora 263601, Uttarakhand, India*
[4] *Micro-Nano Optics Center, CSIR-Central Scientific Instruments Organization, sector 30-C, Chandigarh 160030, India*

*Corresponding author: krakseshsingh.phy@iitbhu.ac.in

**We report the reconstruction of a transparency image in Ghost diffraction scheme using a statistical optics approach. This is implemented by using a static diffuser rather than a pseudo thermal light source with a rotating diffuser. The experimental implementation makes use of spatial ergodicity and spatial stationarity for spatially distributed random fields. A strategy to realize the Ghost diffraction scheme through spatial intensity correlation with a phase retrieval algorithm permits reconstruction of the transparency.**

Image recovery elicited from the correlation of a stochastic light has been the subject of significant research interests in the classical and quantum domains [1–3]. Over the past few years, a plethora of attention has been shown to recover the image using the intensity correlation between two light fields, and techniques such as ghost diffraction (GD) and ghost imaging (GI) have emerged [4–8]. Compared to conventional imaging techniques, the GD & GI uses a non-spatially bucket detector to collect light originating from a transparency either in reflection or transmission geometry. In these techniques, one light field interacts with the transparency and detected by a bucket detector. The second light field which does not interact with the object, directly propagates to the spatially resolved detector having array of pixels. A spatial structure and diffraction pattern of the transparency image can be retrieved by correlating the information measured by the bucket with the information recorded by the array detector in the second arm which never interacts with the transparency. The GD and GI were initially demonstrated with the entangled photons generated in spontaneous parametric down-conversion [9,10]. It was demonstrated later that quantum entangled sources are not necessary and realizations of GD and GI are also possible with the classical correlated light[11] as well as thermal light [12,13]. Since then, significant attention has been attributed to these techniques with classical light for applications in remote sensing [14], biomedical optics [15], microscopy [16], temporal imaging [17], averaged speckle patterns [18], etc.

However, major interests in the GD and GI are limited to recovery of only the modulus square of the Fourier spectrum [13,19,20], without a phase recovery except for some investigations. A modified Young's interferometer was employed to measure the field correlation in the GD and to recover the phase information[21]. A few ghost schemes have also been developed with specific attention to phase recovery[22–24]. Recently, phase recovery in the GD has been demonstrated using the interference of coherent waves using the spatially fluctuating field. This technique uses a combination of Mach-Zehnder interferometers, i.e., inner and outer interferometer assembly [25]. Experimental implementations on the GD & GI usually employ rotating ground glass (RGG) to mimic pseudo thermal light source. Such experimental implementation uses temporal averaging as a substitute for ensemble averaging on the premise of temporal stationarity and temporal ergodicity.

On the other hand, a spatial statistical regime is equally important and needs attention in a situation where temporal averaging is not possible. For instance, spatial statistics of the randomly scattered coherent field, such as laser speckle, provide useful statistical information. Moreover, it permits the replacement of the ensemble average by the spatial average under consideration of spatial stationarity and spatial ergodicity [26].

In this work, we propose and demonstrate the GD scheme with a static ground glass (SGG) rather than RGG. The idea is to examine the reconstruction of a transparency image using basic principle of the GD with the spatial statistical optics approach. Moreover, a phase retrieval algorithm is combined with the results of the GD to retrieve transmission function of a planar object. This strategy helps to recover the transparency information using spatial intensity correlation between the random fields recorded by the bucket detector and array of pixels in the GD scheme.

Experimentation on the proposed scheme is carried out, and experimental results are compared with simulation results. A good agreement between the experiment and simulation supports the use of spatial statistical optics approach for Ghost Diffraction studies. The detailed theoretical explanation, method, corresponding experimental and simulation results are presented below.

A comparison of the GD setup with RGG, and our experimental geometry of the proposed technique with the SGG is shown in Fig. 1. Consider two identical copies of the light field distribution at a source plane, say z=0. Let $E(r)$ denote the optical field at a particular instant of time, i.e., a single realization of the RGG. A beam splitter (BS) is used to create two copies of the optical field. Furthermore, at the plane z=0, one of the two copies illuminates the transparency $\tau(r)$ and propagates to the observation plane. Propagation of the optical field at an observation plane $z=d$ is represented as

$$E_1(u) = \int \tau(r) E(r) G(u-r) dr \quad (1)$$

$$E_2(u) = \int E(r) G(u-r) dr \quad (2)$$

where $G(u) = \exp(iku^2/2d)$ represents the Fresnel propagator, $u$ is the position vector at the observation plane and $k = 2\pi/\lambda$ with wavelength of light $\lambda$. The transparency information is encoded into the spatial correlation functions of the intensities $\langle I_1(u_1) I_2(u_2) \rangle$, where $I_p(u) = |E_p(u)|^2$ is the intensity recorded at the observation plane at a location $u$ in the $p$th beam, where $p = 1,2$.

The correlation of the intensity fluctuations is represented as

$$|g^2(u_1, u_2)| = \langle \Delta I_1(u_1) \Delta I_2(u_2) \rangle \quad (3)$$

where $\Delta I_p(u) = I_p(u) - \langle I_p(u) \rangle$ is the fluctuation of intensity at a location u over its mean value, and $\langle \ \rangle$ represents the ensemble average.

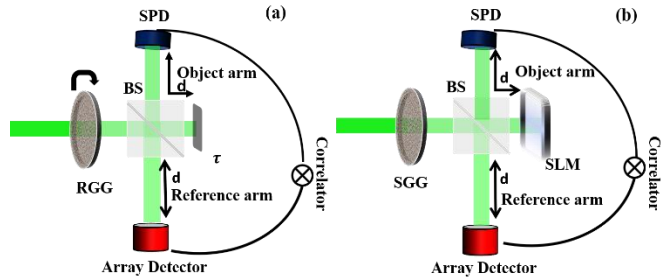

Fig.1. Conceptual representations of a GD scheme with RGG and our proposed GD scheme (a) ghost diffraction scheme with RGG: rotating ground glass, SPD: Single Pixel Detector, transparency-reflective type. (b) Proposed ghost diffraction scheme with SGG: Static Ground Glass, BS: Beam Splitter

Usually, the RGG generates temporally fluctuating speckle patterns for realization of the ensemble averaging as shown in Fig. 1(a). Here, we present realization of the ensemble average by the spatial averaging [26], and two replicas of the speckle pattern are utilized. Therefore, RGG is replaced by a static ground glass (SGG) as shown in Fig.1 (b). Utilizing the spatial stationarity at the observation plane and considering $u_1 = u$ and $u_2 = u + \Delta u$, the correlation of the intensity fluctuations is represented as,

$$g^2(\Delta u) = \left| \tau\left(k \frac{\Delta u}{d}\right) \right| \quad (4)$$

In arriving at Eq. (4), we make use of $\int \exp\left[\frac{-ik}{d}(r_2 - r_1).u\right] du = \delta(r_2 - r_1)$ while implementing the spatial averaging at the observation plane.

Eq. (4) states that the correlation between intensity fluctuations reconstructs the amplitude of the Fourier spectrum of the transparency. Moreover, Fourier spectrum depends only on the difference between two coordinates, i.e., $(u_2 - u_1)$. The spatial distribution of the Fourier spectrum is obtained by spatially varying the detection point in the reference arm and keeping the detection position of the transparency arm at $u_1 = 0$. However, the phase distribution of the Fourier spectrum is lost in Eq. (4) and this obstructs recovery of the transparency $\tau(r)$. To overcome this issue, we feed the correlation of the intensity fluctuations with the Fienup-type iterative phase retrieval algorithm [27–29]. The phase retrieval technique is comprised of hybrid input-output and error reduction algorithms. This strategy helps to completely retrieve the transparency function.

To demonstrate the proposed technique, we simulate the experimental scheme as shown in Fig. 1(b) and confirm it experimentally. Experimental setup of the proposed technique is shown in Fig, 2. A monochromatic laser beam (532nm) is spatially filtered and collimated with the help of a collimating lens(L). This collimated beam of diameter 4.8 mm illuminates a SGG, and the randomly scattered field splits into two parts by a BS. The light transmitted from the BS is dumped. The reflected beam goes to a spatial light modulator TNLC-SLM (LC-R720 model manufactured by HOLOEYE) to insert transparency $\tau(r)$ into the light. The beam reflected from SLM goes back to the BS, and travels to a detector where we use a single pixel of the detector. A polarizer (P) with axis horizontal, is placed right before the detector to make the uniform polarization in the recorded speckle pattern. The detection plane is located at a distance d= 150 mm from the SLM. In the second case, the SLM is switched off and the speckle pattern reflected by the SLM travels back to the detector plane and is recorded by an array of pixels of the charged coupled device (CCD). The CCD camera is with a pixel resolution 2200×2752 and pixel size 4.54 $\mu m$ (Procilica, GT-2750).

Detailed procedure to evaluate the spatial intensity correlation in the GD scheme with the spatial statistical optics is explained in Fig. 3. A random field reflected by the transparency is recorded by detector and only a single pixel data is used as shown in Fig. 3(a). On the other hand, the second random field is recorded by array of pixels with the SLM off condition, i.e., without any transparency function. Therefore, stage of ensemble averaging is shifted from the rotating diffuser to the positional scanning of the spatially fluctuating random fields at the observation plane.

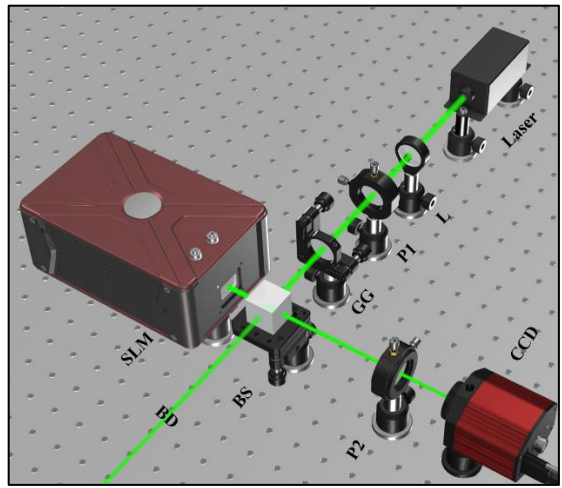

Fig.2. Schematic of the experimental setup of the proposed technique, L: lens, P$_n$(n=1,2): Polarizers, GG: Ground glass, SLM: Spatial Light Modulator, BS: Beam splitter, CCD: Charge-Coupled device. BD-Beam Dumped

Taking a portion of the speckle pattern as a matrix $I_p(u_x, u_y)$ which represents one realization of the randomly scattered field as shown in Fig. 3. Here x and y are pixel spatial coordinates and may take values up to 300×300 pixels. For the field with transparency, we select only a central single pixel out of the 300x300 window, i.e., pixel (151,151) as shown in Fig. 3 (a). The array of pixels 300×300 presents speckle without transparency as shown in Fig. 3(b). A red rectangle with an arrow in Fig. 3(a) and 3(b) represents spatial scanning of the random fields. The cross-covariance of the intensity pattern is obtained by correlating $\Delta I_1(0,0) \Delta I_2(u_x, u_y)$ for different scanning positions as marked in Fig. 3(a) and 3(b), and the process of scanning is represented as $\sum_{m=1}^{M} \left( \Delta I^m_1(0,0) \Delta I^m_2(u_x, u_y) \right) / M$. Here M represents number of different windows of the matrix $I_p^m(u_x, u_y)$ and produced by the pixel-by-pixel movement of the matrix $I_p(u_x, u_y)$ over the speckle. We have used a speckle pattern of size 1000×1000 pixels and 2D scanning of $I_p(u_x, u_y)$ over the speckle patterns provides 700×700 different realizations. The spatial intensity correlation results for simulation and experiment for a transparency, i.e., an annular aperture, are presented in Fig. 3 (c) and 3(d) respectively. A random field is recorded at d and intensity correlation permits realization of a lensless Fourier transform even in the Fresnel domain.

Furthermore, recovery of two different transparencies, namely annular ring and sinusoidal grating from the intensity correlation is demonstrated. For simulation, we evaluate the Fourier pattern from the cross-correlation of intensity fluctuations as described earlier. A phase retrieval algorithm is applied to the far-field pattern to reconstruct the transparency.

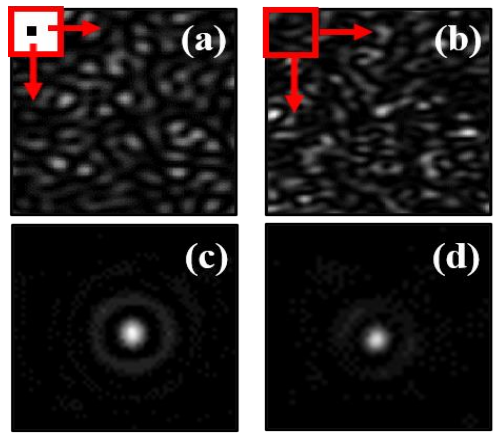

Fig.3 Simulated Speckles: (a) only a single pixel of random field with transparency and scanning of single pixel in the space (b) window of the reference speckle and its scanning (c) and (d): Far-field pattern for annular aperture by simulation and experiment respectively.

A phase retrieval algorithm (Fienup-type algorithm) is applied to the far-field pattern, using constraints that the $\tau(r)$ is real and positive. Moreover, we have also used a loose support constraint (the support is the set of points over which the object function is nonzero). Before applying the phase retrieval algorithm, a 2D Tucky window was used on the measured far field, to remove the noisy sections in intensity correlation function. A standard version of the phase retrieval algorithm is implemented in MATLAB software consisting of a hybrid input-output (HIO) algorithm with a fixed β value. An Error-Reduction algorithm is used to reduce the remaining noise from the retrieved image. β is the feedback parameter that controls the convergence properties of the HIO algorithm. [29].

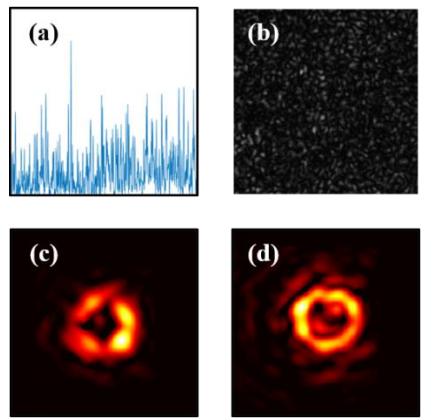

Fig. 4. (a)1-D line profile for single-pixel used from object (Ring) experimental speckle (b) Experimentally recorded speckle without object information. (c) and (d) indicate experimental and simulation reconstructed object transparency $\tau(r)$

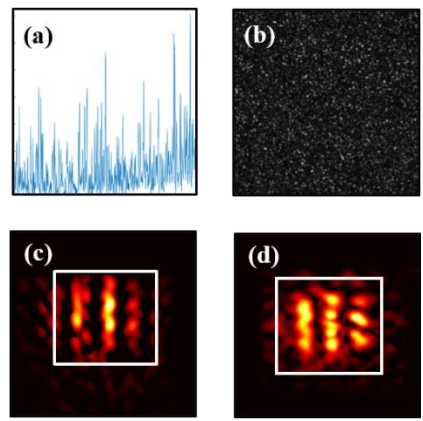

Fig.5.(a)1-D line profile for single-pixel used from Object (sinusoidal grating) (b) Experimentally recorded speckle without object information. (c) and (d) indicate experimental and simulation reconstructed object transparency $\tau(r)$

Experimentally recorded random patterns are shown in Fig. 4(a) and 4(b). Fig.4(a) shows a 1D pattern detected by a single pixel at different spatial points for a binary transmittance function 'ring'. A 2D reference random field without transparency is shown in Fig. 4(b). Reconstruction of the transparency from the intensity correlation of these two fields is shown in Fig. 4(c). Reconstruction of the 'ring' from the simulated random fields is also demonstrated and shown in Fig. 4(d). We have also evaluated the reconstruction of a gradually changing transparency such as a sinusoidal function. The size of this transparency is restricted by an aperture of size 4.8 mm. Reconstruction of a sinusoidal grating transparency is shown in Fig. 5. Fig 5(a) and 5(b) are 1D and 2D random fields and reconstruction of the transparency is shown in Fig. 5(c) and 5(d) for experimental and simulated cases respectively.

In conclusion, we have proposed and experimentally demonstrated a new ghost diffraction scheme by exploiting the correlation features of the spatially varying random fields. Based on the theory mentioned, the technique is capable of reconstructing the object by utilizing the spatial points in the random pattern. The novelty of the technique lies in the significant use of spatial averaging over temporal averaging as used in conventional ghost diffraction scheme. The simulation and experimental results are presented for two different objects. The technique can be extended for the polarimetric ghost objects, which opens a wide range of applications in microscopy, imaging and encryption.

**Funding.** This work is supported by Science and Engineering Research Board (SERB) India- CORE/2019/000026 and Council of Scientific and Industrial Research (CSIR), India- Grant No 80 (0092) /20/EMR-II

**Acknowledgment.** Manisha acknowledges fellowship from the IIT (BHU). Vipin Tiwari would like to acknowledge support from DST-INSPIRE (IF-170861).

**Disclosures.** The authors declare no conflicts of interest.

## References

1. R. H. Brown and R. Q. Twiss, Nature 177, 27 (1956).
2. J. Rosen, V. Anand, M. R. Rai, S. Mukherjee, and A. Bulbul, Appl. Sci. 9,605 (2019).
3. A. Gatti, E. Brambilla, M. Bache, and L. A. Lugiato, Phys. Rev. A 70,013802 (2004).
4. M. J. Padgett and R. W. Boyd, Philos. Trans. R. Soc. A Math. Phys. Eng. Sci. 375,20160233 (2017).
5. P. A. Moreau, E. Toninelli, T. Gregory, and M. J. Padgett, Laser Photonics Rev. 12,1700143 (2018).
6. J. H. Shapiro and R. W. Boyd, Quantum Inf. Process. 11, 949 (2012).
7. C. Zhao, W. Gong, M. Chen, E. Li, H. Wang, W. Xu, and S. Han, Appl. Phys. Lett. 101,141123 (2012).
8. M. Padgett, R. Aspden, G. Gibson, M. Edgar, and G. Spalding, Opt. Photonics News 27, 38 (2016).
9. T. B. Pittman, Y. H. Shih, D. V. Strekalov, and A. V. Sergienko, Phys. Rev. A 52,R3429 (1995).
10. D. V. Strekalov, A. V. Sergienko, D. N. Klyshko, and Y. H. Shih, Phys. Rev. Lett. 74, 3600 (1995).
11. R. S. Bennink, S. J. Bentley, and R. W. Boyd, Phys. Rev. Lett. 89,113601 (2002).
12. A. Gatti, E. Brambilla, M. Bache, and L. A. Lugiato, Phys. Rev. Lett. 93,093602 (2004).
13. F. Ferri, D. Magatti, A. Gatti, M. Bache, E. Brambilla, and L. A. Lugiato, Phys. Rev. Lett. 94,183602 (2005).
14. N. D. Hardy and J. H. Shapiro, Phys. Rev. A 84,063824 (2011).
15. T. Shirai, T. Setälä, and A. T. Friberg, J. Opt. Soc. Am. B 27, 2549 (2010).
16. Z. Sun, F. Tuitje, and C. Spielmann, Opt. Express 27, 33652 (2019).
17. D. Faccio, Nat. Photonics 10, 150 (2016).
18. P. Zerom, Z. Shi, M. N. O'Sullivan, K. W. C. Chan, M. Krogstad, J. H. Shapiro, and R. W. Boyd, Phys. Rev. A 86,063817 (2012).
19. D. Zhang, Y.-H. Zhai, L.-A. Wu, and X.-H. Chen, Opt. Lett. 30, 2354 (2005).
20. A. Valencia, G. Scarcelli, M. D'Angelo, and Y. Shih, Phys. Rev. Lett. 94,063601 (2005).
21. R. Borghi, F. Gori, and M. Santarsiero, Phys. Rev. Lett. 96,183901 (2006).
22. P. Clemente, V. Durán, E. Tajahuerce, V. Torres-Company, and J. Lancis, Phys. Rev. A 86,041803 (2012).
23. T. Shirai, T. Setälä, and A. T. Friberg, Phys. Rev. A 84,041801 (2011).
24. D. J. Zhang, Q. Tang, T. F. Wu, H. C. Qiu, D. Q. Xu, H. G. Li, H. B. Wang, J. Xiong, and K. Wang, Appl. Phys. Lett. 104,121113 (2014).
25. R. V. Vinu, Z. Chen, R. K. Singh, and J. Pu, Optica 7, 1697 (2020).
26. M. Takeda, W. Wang, D. N. Naik, and R. K. Singh, Opt. Rev. 21, 849 (2014).
27. J. R. Fienup, Opt. Lett. 3, 27 (1978).
28. J. R. Fienup, Appl. Opt. 21, 2758 (1982).
29. B. Das, N. S. Bisht, R. V. Vinu, and R. K. Singh, Appl. Opt. 56, 4591 (2017).